\documentclass[a4paper,11pt]{article}
\pdfoutput=1 % if your are submitting a pdflatex (i.e. if you have
\usepackage{jinstpub} % for details on the use of the package, please
\usepackage{siunitx}
\usepackage{todonotes}
\usepackage{subfig}
\usepackage[export]{adjustbox}
\usepackage{placeins}
\usepackage{multirow}
\usepackage[14pt]{extsizes} % Правильно меняет размер текcта
\usepackage[utf8]{inputenc}	%Кодировка utf8	
\usepackage[T2A]{fontenc} % поддержка кириллицы
\usepackage{pdfpages}
\begin{document}
\title{Note on Intrinsic Resolution in Liquid Organic Scintillators}
\author[a]{O.\ Smirnov }

\affiliation[a]{Joint  Institute  for  Nuclear  Research,  141980,  Dubna,  Russian Federation}

% e-mail addresses: only for the corresponding author
\emailAdd{osmirnov@jinr.ru}

\maketitle
\begin{abstract}

Fluctuations in photon production in scintillator could contribute to the total energy resolution of a scintillation detector. This contribution, called intrinsic resolution (IR), is one of the factors limiting the total energy resolution in detectors based on liquid organic scintillators (LSs). There are very few experimental measurements of the IR in LSs available, and the underlying physics is not completely understood. We propose a phenomenological description of IR of LSs and systematize the data available using a single universal parameter that characterizes IR. We show that all experimental data within the model demonstrate the presence of extra smearing of energy resolution due to IR with a typical value of  $\simeq2$\% at 1-MeV energy release. The model can be used to simulate the effect of IR in LS-based detectors.
\end{abstract}

\section{Introduction}

The statistical properties of a scintillation detector response were studied by Breitenberger~\cite{Breitenberger} and independently by Wright~\cite{Wright}. The relative variance $v(Q)$ of a PMT signal with a mean number of photoelectrons (p.e.) $Q$, registered in a scintillation event with an average number of produced photons $\overline{N_{ph}}$  can be expressed as:

\begin{equation}
v(Q)\equiv\left(\frac{\sigma_{Q}}{Q}\right)^{2}=\frac{1+v_{1}}{Q}+v(p)+(1+v(p))\left[v(N_{ph})-\frac{1}{\overline{N_{ph}}}\right].
\label{Birks}
\end{equation}

The parameter $v(N_{ph})$, of primary interest for this note, is the relative variance of photon production. For normally distributed production of photons, the term $v(N_{ph})=\frac{1}{\overline{N_{ph}}}$
and the last term vanish. 

The parameter $v(p)$ characterizes the relative variance of photon detection efficiency over all possible paths from the origin to detection, and the parameter $v_{1}$ is the relative variance of a single photoelectron response (SER) of a PMT. Sometimes, the excess noise factor (ENF) is used instead of the $1+v_{1}$ term defined as $ENF=1+v_{1}$. 

The formula (\ref{Birks}) was derived for a single PMT. In the case of a detector with many PMTs, a corresponding expression can be obtained with similar terms~(see e.g.~\cite{MyResolutions,BrxPhaseI}).

The origin of the excess over normal variations (i.e., $\frac{1}{N_{ph}}$) during photon production was widely discussed for non-organic scintillators (see, e.g.,~\cite{Moszynski}). This excess is usually attributed to scintillator non-linearities in the energy response, which are, in turn, the result of Landau fluctuations and secondary electron production ($\delta$-rays). The mechanism is not confirmed by simulations for liquid organic scintillators~\cite{Intrinsic}. The authors refer to the fluctuation in the energy transfer process, which is much less efficient compared to the one in non-organic crystals.

Note that, in contrast to all other parameters, the $v(N_{ph})$ parameter is not universal as it is defined for a particular value of $\overline {N_{ph}}$ corresponding to the given energy of the particle. In this paper, we would like to derive a universal parameter that will allow comparing experimental results at various energies and photoelectron yields.

\section{Phenomenological Model}

In general, the parameter $v(N_{ph})$ should be measured for the energy of interest. Nevertheless, some simple considerations can be used to understand the energy dependence of this parameter. First, let us note that the total amount of photons produced in an event with the energy release E could be expressed as:
\begin{equation}
N_{ph}=LY_{ph}\cdot E\cdot f_{NL}(E,...),   
\end{equation}
where $LY$ is the relative photon (light) yield per MeV and $f_{NL}(E,...)$ is the function describing non-linearity of the energy response, the notation "..." is for a set of possible parameters, which are of no importance here and will be skipped in notations.  The light yield and non-linearity factor depend on the type of interacting particle. In our study, we consider only electron interactions. With low light intensities at a single PMT of the detector and linearity of electronics, one can expect a linear dependence of the average amount of the detected photoelectrons $Q$ on the number of average photons produced in the interaction: $Q=c_Q\cdot N_{ph}$. From here on, we will discuss the average values of the involved variables, if otherwise is not explicitly stated. The total amount of the photoelectrons collected is:
\begin{equation}
Q=LY \cdot E\cdot f_{NL}(E),   
\end{equation}
where $LY\equiv LY_{ph}\cdot c_Q$  is the relative photoelectron yield per MeV.

Let us introduce the relative variance of the number of produced photons $v(N_{ph}(1))$ at a 1-MeV energy release for electrons~\footnote{Energy release fluctuations for $\gamma$'s are different because of multiple Compton scattering and scintillation non-linearities, that is why they are not discussed here. We also do not discuss scintillations caused by $\alpha$-particles.}. The corresponding charge deposit can be expressed as $Q_{1}=LY \cdot f_{NL}(1)$. The $Q_1$ value coincides with the relative photoelectron yield of the detector if the non-linearity factor is normalized in a way to provide unity at 1 MeV ($f_{NL}(E)\equiv1$). We will keep the notation $Q_1$ in order to underline the importance of the physical amount of photoelectrons in the study; the photoelectron yield could be generally defined in a way that complicates its usage for the calculations.

If the variance of the number of photons produced for a 1-MeV particle is $\sigma_{N_{ph}}^2(1)$, then: 

\begin{equation}
\sigma_{N_{ph}}^2(E)\simeq \frac {N_{ph}(E)}{N_{ph}(1)}\sigma_{N_{ph}}^2(1),
\end{equation}

where the approximate equality sign is used to underline a possible non-linearity of intrinsic smearing with respect to energy. If the hypothesis of fluctuation in the energy transfer process is valid, the equality should be quite precise. The relative variance of the number of the photons produced is:

\begin{equation}
\begin{split}
v(N_{ph}(E))=\frac {\sigma_{N_{ph}}^2(E)}{N_{ph}(E)^2}=\frac {\sigma_{N_{ph}}^2(1)}{N_{ph}(E) \cdot N_{ph}(1)}=\frac {N_{ph}(1)}{N_{ph}(E)} \frac {\sigma_{N_{ph}}^2(1)}{N_{ph}^2(1)}\equiv
\\
\equiv\frac {N_{ph}(1)}{N_{ph}(E)}\frac{N_{ph}(1)+\delta_{N_{ph}}^2(1)}{N_{ph}^2(1)}
=\frac {1}{N_{ph}(E)}+\frac{N_{ph}(1)}{N_{ph}(E)}v_{1}^{int}.
\end{split}
\label{varN_}
\end{equation}

Here, we introduced $\delta_{N_{ph}}^2(1)$, an "extra" variance with respect to the pure Poisson variance at 1-MeV energy release $N_{ph}(1)$, and defined the relative extra variance at 1 MeV as $v_{1}^{int}\equiv\frac{\delta_{N_{ph}}^2(1)}{N_{ph}^2(1)}$.

It is convenient to rewrite (\ref{varN_}) through the charge $Q$ in order to use it in (\ref{Birks}):

\begin{equation}
\begin{split}
v(N_{ph}(E))-\frac{1}{N_{ph}(E)}=\frac {f_{NL}(1)\cdot LY_{ph}\cdot 1}{f_{NL}(E)\cdot LY_{ph}\cdot E}v_{1}^{int}=
\\
=\frac {LY\cdot f_{NL}(1)}{Q} v_{1}^{int}\equiv\frac {Q_1}{Q} v_{1}^{int},  
\end{split}
\label{varNph}
\end{equation}

where $Q_1$ is the charge corresponding to a 1-MeV energy release for electrons. Thus, the term $\left[v(N_{ph})-\frac{1}{\overline{N_{ph}}}\right]$ in formula (\ref{Birks}) can be substituted by:

\begin{equation}
    v_{int}\equiv\frac {Q_1}{Q} v_{1}^{int}.
    \label{eq:v_int}
\end{equation}

We get rid of $N_{ph}$ and its variance in the expression, introducing the only universal parameter $v_1^{int}$. The value of the parameter $Q_1$ depends on LS properties and detector characteristics. The formula is applicable to any type of particle, but the values of $v(p)$ and $v_{1}^{int}$ can differ for electrons and gammas for the same charge collected. Note that $v(p)$ can be neglected due to the typically small value compared to unity; e.g., for a 10\% non-uniformity, we have $v(p)=0.01 \ll 1$.

As one can see from~(\ref{eq:v_int}), in fact, the extra variance due to the intrinsic resolution contributes to the "statistical" term $\frac{1+v_1}{Q}$ as it has the same energy dependence $\frac{1}{Q}$. Basically, one can use the "efficient" $v_1$ (or ENF) parameter in the form $v_1^{eff}=v_1+Q_1 v_1^{int}$ ($\text{ENF}_{eff}=1+v_1+Q_1 v_1^{int}$). 

So, the intrinsic resolution effect mimics statistical smearing. It is more pronounced for detectors with larger photoelectron yields. Indeed, for a 1-MeV energy release, the contribution of  $v_1^{int}$ should be compared to $\frac{1+v_1^{Det}}{Q_1}$ — the bigger the $Q_1$ parameter, the bigger the relative contribution of IR.

Taking into account the low $v_1^{int}$ values of a few units of $10^{-4}$ (see the following discussion and final table~\ref{Table:AllData}), it is clear that experiments with a high photoelectron yield are more suitable for the $v_1^{int}$ measurement. In fact, for the expected intrinsic resolution of 2\% at 1 MeV and the photoelectron yield of about 100 p.e./MeV, the extra term is $10^{2}0.02^2=0.04$, a rather low one compared to the $1+v_1$ values. In this case, one should provide at least a $\simeq 1$\% precision measurement of the statistical term in order to separate the $Q_1 v_1^{int}$ contribution. 

\section{Direct Measurements of Intrinsic Resolution in Liquid Organic Scintillators}

Recently, two measurements have been performed using Compton spectrometers and LAB-based scintillators. In paper~\cite{Formozov}, these were the measurements with the $^{137}$Cs monoenergetic $\gamma$-source. The reported value characterizing intrinsic energy resolution is $v_{int}(50\text{ keV})=0.02\pm0.005$. Since
$v_{int}=\frac{Q_1}{Q}v_1^{int}=\frac{LY\cdot 1}{LY\cdot 0.05\cdot f_{NL}(E,kB)}v_1^{int}\simeq \frac{20}{0.9} v_1^{int}$, we obtain $v_1^{int}=(0.0011\pm0.0003)$ corresponding to a $(3.3\pm0.4)$\% intrinsic resolution at 1 MeV. The value $f_{NL}(k_B,E_e=0.05\text{ MeV})\simeq 0.9$ is derived from figure 3 of~\cite{Formozov}.

In the paper~\cite{Intrinsic}, the value of $(1.83\pm0.06)$\% is reported for monoenergetic 0.976-~MeV electrons. The formula used for the analysis contains the sum of terms attributed to all possible contributions, and the term responsible for IR has a simple form, $\delta_{int}^2$. If we compare the formula with (\ref{Birks}), we can see that the factor $(1+v(p))$ is missing (note that in~\cite{Intrinsic} $v(p)$ is called $\delta_{uniformity}^2$), but due to the small value of $\delta_{uniformity}^2=(0.89\times10^{-2})^2$,  factor can be neglected. We can derive $v_1^{int}=(3.3\pm0.2)\times10^{-4}$ corresponding to a $1.82\pm0.06$\% intrinsic resolution at 1 MeV.

\section{Intrinsic Resolution in the Borexino Analysis}

In contrast with the direct measurements performed with LAB-based LSs discussed in the previous section, the Borexino detector used LS on the base of pseudocumene (PC). We are interested in the analysis at low energies, where the term responsible for IR appears in analytical expressions for energy resolution. An important point in the Borexino analyses is the energy estimator $N$, namely the amount of triggered PMTs in an event\footnote{PMT occupancy, defined as the fraction of detected events  $\frac{N_{detected}}{N_{triggers}}$ in a series of trials, is genetically the same estimator if applied to a single event in a many-PMT detector. The number of trials in this case coincides with the amount of working PMTs.} \cite{Nature,pp,eDecay,mm,BrxSpectroscopy}. The choice of this estimator instead of the above-mentioned simple p.e. counting is justified mainly by the better energy resolution achievable in the high-segmented detector in the sub-MeV region. We would like to note another advantage of using this energy estimator: it doesn't depend on the linearity of the PMT electronics channel as well as on the calibration precision of the individual PMTs of the detector (conversion of a rough ADC measurement into the amount of true photoelectrons), which is, in turn, a non-trivial problem mainly due to the large amount of PMTs in this kind of LS detector.

The broadening of resolution with respect to the models with no intrinsic resolution was noticeable, so an extra term was included in the Borexino analytical model. In order to convert the Borexino parameter into the one from our phenomenological model, let us consider the smearing of the $N$ energy estimator occurring due to intrinsic resolution.

First, we note that the variance of the average number of the detected photoelectrons $\mu$ with each PMT of the detector is:

\begin{equation}
v(\mu(E))=\frac{\mu_1}{\mu}v_1^{int},
\label{eq:varMu_vN1}
\end{equation}

where $\mu$ is the average amount of p.e. detected in an event with the energy E and $\mu_1$ corresponds to a 1-MeV energy release.

We are interested in the extra term in the expression for the signal variance due to the intrinsic smearing. Let us consider the simplest case of a point-like source at an ideal spherical detector's centre. The variance of the number of  triggered PMTs for the events with the fixed average $\mu$ value follows the binomial distribution:

\begin{equation}\sigma_{N}^2= N_{PMT}p_0(1-p_0),
\label{eq:sigmaN0}
\end{equation}

where $N_{PMT}$ is the total number of PMTs of the detector and $p_0$ is the probability of the absence of the signal at a single PMT. We are interested only in the IR effect, so we will not take into consideration other effects, such as the electronics threshold effect or variance of the p.e. detection efficiency through the photocathode and angles of photon incidence. In this case, $p_0$ corresponds to the Poisson probability P(0), and we will use $p_0=e^{-\mu}$ in calculations.

We will need the average values of $p_0$ and $p_0^2$ over all events and the average values of $\mu$ distributed in accordance with the $\rho(\mu)$ distribution for further calculations. Assuming the normal distribution of p.d.f. of the extra smearing $\rho(\mu)$ with the average $\mu_0$ and the relative variance from (\ref{eq:varMu_vN1}),  we obtain for $\overline {p_0}$:

\begin{equation}
\begin{split}
\overline {p_0}=\frac{1}{N_{events}}\sum_i^{N_{events}} e^{-\mu_{i}}\simeq
 \int e^{-\mu}\rho(\mu)d\mu=e^{-\mu_0 (1-\frac{\mu_1}{2}v_1^{int})},
 \label{eq:p0_avg}
\end{split}
\end{equation}

 and for $\overline {p_0^2}$:

\begin{equation}
\begin{split}
\overline {p_0^2}=\frac{1}{N_{events}}\sum_i^{N_{events}} e^{-2\mu_{i}}\simeq
 \int e^{-2\mu}\rho(\mu)d\mu=e^{-2\mu_0 (1-{\mu_1}v_1^{int})}=
 \\
=e^{-2\mu_0+2\mu_0 \mu_1 v_1^{int}}=
e^{-2\mu_0+2\mu_0\frac{\mu_1}{2}v_1^{int}+
 2\mu_0\frac{\mu_1}{2}v_1^{int}}
 = \overline{p_0}^2 e^{\mu_0 \mu_1 v_1^{int}}.
 \label{eq:po_sq_avg}
\end{split}
\end{equation}

We used (\ref{eq:p0_avg}) to obtain (\ref{eq:po_sq_avg}). Also, we replaced summation with integration, assuming a large sample of events. Integration is performed over all possible $\mu$ values, and the integral can be calculated analytically in the case of the normal distribution $\rho(\mu)$.

In order to calculate the variance over a set of events, we write:
\begin{equation}
\begin{split}
\sigma_N^2 = \overline{N^2}-\overline{N}^2 = \overline{N_{PMT}^2 (1-p_0)^2+N_{PMT}p_0(1-p_0)}-\overline{N_{PMT}(1-p_0)}^2=
\\
N_{PMT}^2(1-2\overline{p_0}+\overline{p_0^2})+N_{PMT}(\overline{p_0}-\overline{p_0^2}-N_{PMT}^2(1-2\overline{p_0}+\overline{p_0}^2)=
\\
=N_{PMT}^2(\overline{p_0^2}-\overline{p_0}^2)+N_{PMT}\overline{p_0}(1-\overline{p_0}e^{\mu_0\mu_1 v_1^{int}})=
\\
=N_{PMT}\overline{p_0}(1-\overline{p_0}e^{\mu_0\mu_1 v_1^{int}})+N_{PMT}^2\overline{p_0}^2(e^{\mu_0\mu_1 v_1^{int}}-1).
\end{split}
\label{eq:Var000_vN1}
\end{equation}

The statistical part (the first term in \ref{eq:Var000_vN1}) contains an extra factor $e^{\mu_0\mu_1 v_1^{int}}$ that can be neglected due to the small $v_1^{int}$ value ($\mu_1=0.28$ and $\mu$ is well below $\mu_1$ for the energy range of interest in the Borexino analysis). In this way, we can separate an extra term responsible for intrinsic resolution:
\begin{equation}
\begin{split}
\sigma_{int}^2 = N_{PMT}^2\overline{p_0}^2(e^{\mu_0\mu_1 v_1^{int}}-1).
\end{split}
\label{eq:Sigma_int}
\end{equation}

Equation (12) from the Borexino paper~\cite{BrxSpectroscopy} contains, in its turn, the following term responsible for intrinsic resolution\footnote{The multiplier $10^{-2}$ is missing in the article cited, we put it back here; also, we skip the irrelevant factor $f_{eq}$ as it appears because of the varying number of PMTs, and it keeps the factor $v(N)$ stable with respect to the initial amount of PMTs.}:

\begin{equation}
\sigma_{int}^{2}
=10^{-2}v(N)\cdot N.
\label{eq:SigNSqBrxEq}
\end{equation}

The term responsible for intrinsic resolution in (\ref{eq:Var000_vN1}) is reduced to the one used in the Borexino approximation for $\mu \ll 1$. The approximation is reasonable, especially when taking into account that most of the statistics in the Borexino low-energy analysis are concentrated at the end-point of the $^{14}$C $\beta$-decay, i.e., 156 keV, and the corresponding $\mu$ value at a single PMT of the total of 2000 is $\mu\simeq0.156\frac{551}{2000}=0.04$ p.e.:

\begin{equation}
\begin{split}
    N_{PMT}^2p_0^2(e^{\mu \mu_1 v_1^{int}}-1)\simeq N_{PMT}^2 p_0 \mu \mu_1 v_1^{int}=N_{PMT}N\frac{p_0}{p_1}\mu_1 \mu v_1^{int} \simeq 
\\    
\simeq N_{PMT}N\frac{1-\mu}{\mu}\mu_1 \mu v_1^{int}=N (N_{PMT}\mu_1)v_1^{int}=N Q_1 v_1^{int},
\end{split}
\label{eq:BrxReduction}
\end{equation}

the parameter $v_1^{int}$ is related to the Borexino's $v(N)$ as\footnote{Note that $v(N)$ is used as a constant in the Borexino analysis.}:

\begin{equation}
    v_1^{int}\simeq10^{-2}\frac{v(N)}{Q_1}.
    \label{eq:v1_brx}
\end{equation}

The value of the parameter $v(N)$ reported in~\cite{BrxSpectroscopy} is $11.5\pm1.0$, and thus we can derive $v_1^{int}=(2.2\pm0.2)\times10^{-4}$ using the value of $LY=(551\pm 1)$ p.e./[MeVx2000 PMT] from ~\cite{BrxSpectroscopy} as $Q_1$.
%%, the value $f_{NL}(0.156\text{ MeV})=0.95$ is calculated using the Borexino value of $k_B=0.0109$ cm/MeV. 
This value of $v_1^{int}$ corresponds to the intrinsic resolution of $(1.5\pm0.1)$\% at 1 MeV.

In the earlier Borexino collaboration papers~\cite{pp,mm,eDecay} devoted to the low-energy analysis, the constant term in the form $\sigma_{int}^2$ was used instead of the energy-dependent one. The values of the fit parameter are reported in~\cite{Pablo_PhD}: $\sigma_{int}=1.69\pm0.23$ and $LY=420.1\pm1.4$ p.e./MeV. No systematic study of the parameters is available, so we will use these values as the indicative ones. Similar to the previous case, the value of the parameter corresponds to the effective energy close to the end-point of the $^{14}$C $\beta$-decay. The fit in the early Borexino low-energy studies was performed with the energy estimator not corrected for the varying live PMT amount, and the LY=420.1 p.e./MeV corresponds to the average number of PMTs operating in this period. The parameter $\sigma_{int}$ is linked to our universal parameter $v_1^{int}$ by the relation:
\begin{equation}
\sigma_{int}^2=NQ_1 v_1^{int}.
\label{eq:v1_brx_2}
\end{equation}

The lower limit in the Borexino fit, N=60, corresponds to 165 keV~\cite{pp}. In order to estimate the effective N value, we should estimate the value of the effective energy for the fraction of the $^{14}$C spectrum above the threshold. Apparently, the value is below the end-point of 156 keV. Taking into account that the end-point of the visible $^{14}$C spectrum is smeared up to 230 keV, we can use as an estimate of the average value the scaled low bound of the smeared spectrum: $\simeq 156\frac{165}{230}$ keV, or $\simeq 110$ keV. So, for $E=0.11$ MeV $N=40$, we got $v_1^{int}\simeq(1.7 \pm0.5)\times10^{-4}$.

\section{Discussion of Results}

All the data are compiled in table~\ref{Table:AllData}. As one can see, the data indicate the presence of extra smearing due to intrinsic resolution at the level of $\simeq(1\div3)$\%. The data from the experiments differ, but since scintillator mixtures were not the same, it is difficult to draw further conclusions: the difference in values could be due to LS cocktail components. The two measurements of Borexino are in good agreement, though no systematic uncertainty study is available. The value from the later Borexino paper should be more precise as the parameterization used in the analysis follows the energy dependence of the parameter, the earlier analysis implied an effective value, and the conversion into the universal parameter needs precise estimation of the effective N value at the tail of the $^{14}$C spectrum. The measurement by Formozov~\cite{Formozov} for LAB-based LSs has a higher value compared to the one by Deng et al.~\cite{Intrinsic}, but this could be the result of different LS cocktails as well. Also, we would like to point out the relatively close values for the most precise measurements~\cite{Intrinsic} and~\cite{BrxSpectroscopy}.

In conclusion, we would like to discuss the implications of our study for the JUNO experiment~\cite{JUNO} where energy resolution is crucial for the success of the neutrino mass hierarchy measurement. Energy reconstruction in JUNO is provided by machine learning algorithms~\cite{JUNO_rec}. Information from all PMTs is fed to the algorithms, which should provide the optimal reconstruction at a given energy. The achievable precision of reconstruction is limited by the detector properties and can differ for the $N$ and $Q$ energy estimators, the machine learning algorithm should choose the optimal one. For completeness, let us consider both classes of energy estimators. 

The inclusion of the IR contribution results in a modification of the statistical part of the expression for the energy resolution for the energy estimator based on the total charge collected:

\begin{equation}
    \sigma_{stat}^{2}(Q)=(1+v_1^{Det})Q+Q_1v_{int}^1 Q=(1+v_1^{Det}+Q_1v_{int}^1)Q,
    \label{JUNO_stat}
\end{equation}

and the statistical term $R^{\prime}_{stat}$ of the energy resolution accounting for intrinsic resolution can be expressed through the "no IR" term $R_{stat}$ as $R^{\prime}_{stat}=R_{stat}\cdot\sqrt{\frac{1+v_1^{Det}+Q_1v_{int}^1}{1+v_1^{Det}}}$. Taking into account the expected LY=1345 p.e./MeV (the value is quoted for the detector center~\cite{JUNO2}; for the whole detector, it should be admittedly higher, but no data are still available, so we will use the value as a lower limit for $Q_1$) and $v_1^{Det}=0.48$\footnote{15,000 MCP PMTs and 5,000 dynode PMTs are ordered for JUNO~\cite{JUNO2}. The corresponding ENFs were reported in~\cite{PMTs}: 1.55 and 1.17, respectively. Based on these values, we calculated $v_1^{Det}$ as the weighted value.}, the estimated degradation of the statistical term is above 14\%  ($R^{\prime}_{stat} \geq 1.14 R_{stat}$)  for $v_1^{int}=3.3\times10^{-4}$ . 

In the case of using the energy estimator based on the number of triggered PMTs, the extra term in resolution is quantified by expression~\ref{eq:Var000_vN1}, or its approximation~\ref{eq:BrxReduction}. The extra variance is:

\begin{equation}
    \frac{Q_1}{N} v_1^{int},
\end{equation}

using the low-energy approximation $N\simeq Q$, we got $ \frac{Q_1}{N} v_1^{int}\simeq v_1^{int}$ at 1 MeV, i.e., we have an extra 1.8\% that should be summed up quadratically with the statistical term of the "standard" resolution at 1 MeV for the $N$ energy estimator.

The degradation in energy resolution could be crucial for the success of the JUNO experiment, so further studies are needed. Based on our model, we can indicate the following simple way of including intrinsic resolution into the standard MC description of the detector: the total amount of photons generated for the fixed energy release should be generated using a normal distribution with an energy-dependent extra width corresponding to the intrinsic smearing: $\sigma_{N_{ph}}^2=\frac{1}{N_{ph}}+\frac{N_{ph}(1)}{N_{ph}}v_1^{int}$.
  
\textbf{\begin{table}
\begin{tabular}{|c|c|c|c|c|}
\hline 
Measurement & LS cocktail & rough value & $v_1^{int}$ & $R_{int}(1\text{ MeV})$ \tabularnewline
& & & $\times 10^{-4}$ & \% \tabularnewline
\hline 
\hline 
Formozov~\cite{Formozov} & LAB+1.5 g/l PPO & $v_{int}(50\text{ keV})=$ & $11\pm3$ &$3.3\pm0.4$  \tabularnewline
  &   & $0.02\pm0.005$ &  &   \tabularnewline
\hline 
Deng et al~\cite{Intrinsic} & LAB+2.5 g/l PPO   & $1.83\pm0.06$\%  & $3.3\pm0.2$ & $1.82\pm0.06$\tabularnewline
  &  + 3 mg/l bis-MSB &  @ 0.976 MeV &   & \tabularnewline
\hline 
Borexino~\cite{BrxSpectroscopy} & PS + 1.5 g/l PPO & $v(N)=11.5\pm1$ & $2.2\pm0.2$ & $1.5\pm0.1$\tabularnewline
\hline 
Borexino~\cite{Pablo_PhD} & PS + 1.5 g/l PPO & $\sigma_{int}=1.69\pm0.23$ & $1.7\pm0.5$ & $1.3\pm0.3$\tabularnewline
\hline 
\end{tabular}
\caption{Available data on intrinsic resolution and their conversion to the universal parameter $v_1^{int}$. The last column contains the relative energy resolution at 1 MeV due to intrinsic smearing}
\label{Table:AllData}
\end{table}}

\section{Acknowledgements}
  This work was supported by the Russian Science Foundation, project no. 21-12-00063.            
  The author is grateful to N. Mazarskaya for her valuable help in preparing the manuscript.

\end{document}